\documentclass{article}       

\begin{document}

\newtheorem{definicao}{Definition}
\newtheorem{teorema}{Theorem}
\newtheorem{lema}{Lemma}
\newtheorem{corolario}{Corollary}
\newtheorem{proposicao}{Proposition}
\newtheorem{axioma}{Axiom}
\newtheorem{observacao}{Observation}
\newtheorem{exemplo}{Example}

\title{Individuality, quasi-sets and the double-slit experiment
}


\author{Adonai S. Sant'Anna
}


\date{A. S. Sant'Anna \at
              Department of Mathematics, Federal University of Paran\'a, Curitiba, PR, 81531-990, Brazil \\
              Tel.: +55-41-3085-3564
              \email{adonai@ufpr.br}\\
              ORCID 0000-0003-3425-698X           
}

\date{Published in {\em Quantum Studies: Mathematics and Foundations\/}, doi.org/10.1007/s40509-019-00209-2 (2019)}

\maketitle

\begin{abstract}
Quasi-set theory $\cal Q$ allows us to cope with certain collections of objects where the usual notion of identity is not applicable, in the sense that $x = x$ is not a formula, if $x$ is an arbitrary term. $\cal Q$ was partially motivated by the problem of non-individuality in quantum mechanics. In this paper I discuss the range of explanatory power of $\cal Q$ for quantum phenomena which demand some notion of indistinguishability among quantum objects. My main focus is on the double-slit experiment, a major physical phenomenon which was never modeled from a quasi-set-theoretic point of view. The double-slit experiment strongly motivates the concept of degrees of indistinguishability within a field-theoretic approach, and that notion is simply missing in $\cal Q$. Nevertheless, other physical situations may eventually demand a revision on quasi-set theory axioms, if someone intends to use it in the quantum realm for the purpose of a clear discussion about non-individuality. I use this opportunity to suggest another way to cope with identity in quantum theories.

\end{abstract}


\section{Introduction}\label{intro}

Since its early days, quantum mechanics has motivated a vast amount of unexpected explanations for extraordinary phenomena \cite{Ghirardi-07}, multiple interpretations of mathematical models and physical phenomena \cite{Holland-93} \cite{Byrne-13} \cite{Bacciagaluppi-99} (among many others), several Gedanken experiments \cite{Zeilinger-99}, and many philosophical debates as well \cite{Jammer-74}. Eventually some philosophical perspectives seem to be strong enough to motivate new mathematical approaches. That was what happened with Heinz Post's views concerning (non)individuality in quantum physics \cite{Post-63}. Such a philosophical paper by Post provided a strong case for D\'ecio Krause's Quasi-Set Theory \cite{Krause-92} \cite{French-06}. According to Post, the non-individuality of quantum particles should be ascribed {\em right at the start\/}, as we further discuss in the next sections. And that was what Krause and collaborators did: a mathematical framework inspired on Post's ideas. Nevertheless, Post's ideas concerning the problem of identity in quantum theories were published five decades ago. And a lot has happened in the study of quantum phenomena since then. One of the novelties regards possible violations of the Symmetrization Postulate \cite{Angelis-96} \cite{Modugno-98}. And since the Symmetrization Postulate plays a major role for justifying the use of quasi-set theory in quantum physics \cite{Domenech-08} \cite{Domenech-10} \cite{French-06}, maybe it is time to think again about the relationships between the usual notions of indistinguishability in quantum theories and the quasi-set-theoretic approach for those theories. After all, quasi-set theory is grounded on the belief that quantum indistinguishability entails some sort of non-individuality among quantum objects. And that is a conclusion that may be precipitated.

In order to avoid possible misunderstandings about the contents of this paper, here follows a brief warning. One thing is a mathematical framework used for understanding and modeling physical phenomena. Another situation is the philosophical discussions concerning possible interpretations of formal frameworks. Since my goal here is closely connected to the problem of non-individuality in quantum theories, a few words must be said about terms like ``individuality'', ``non-individuality'', ``identity'', and alike. Many philosophical discussions about individuality (and eventually non-individuality) were motivated by mathematical descriptions and even physical phenomena concerning the quantum realm. And I believe it is safe to say that those discussions were strongly motivated by some notion of {\em indistinguishability\/} commonly used in theoretical physics. For details see \cite{French-06}. Well, if that is true, we need to identify those situations where any notion of {\em indistinguishability\/} plays any important role in physics. Quantum mechanics and non-relativistic quantum field theories are well known examples of physical theories where that happens.

Quantum statistics, for example, demand some radical notion of indistinguishability among particles. That happens because permutations of quantum particles among quantum states are deemed as not observable (even in principle). That corresponds to the Symmetrization Postulate (SP), which is stated here in an intuitive way (as it usually happens in specialized literature about physics). And such an intuitive statement for the SP is, by itself, quite confusing. After all, if permutations are not observable, how can we even know there was any permutation at all? Although unicorns are not observable, that fact does not legitimate any serious study about a legendary horse with a single straight horn projecting from its forehead. If permutations are not observable, does that mean permutations are not possible? What is the difference between one statement (non observability) and the other (impossibility)? One could argue that SP is an invariance principle from physics. But invariance principles are usually stated as valid for certain physical systems $S$ in order to study those properties which are not invariant within subsystems of $S$. And that is not the case of SP. Quasi-set theory, on the other hand, allows us to rewrite SP within a specific and clear formal language. But even in that case, we can see something rather important is still missing. We go back to that point in Section 3.

Another example is the spectral lines of elements. In order to accommodate the observed spectral lines of the Helium atom, for example, we need to take into account some sort of radical indiscernibility between its protons. Both situations (quantum statistics and the Helium atom) are examined in \cite{Krause-99}, where quasi-set theory is successfully employed from a logical point of view.

Nevertheless, there are other situations where indistinguishability plays a fundamental role. One of them is the double-slit experiment (Young interferometer), discussed in the next Section. Since there is no gas of trajectories and since trajectories are no physical components of atoms, that situation deserves a rather special attention. The double-slit experiment strongly suggests a close relationship between indistinguishability of photon trajectories (which is equivalent to indistinguishability between photon sources) and coherence (between fields). That was precisely demonstrated by Mandel \cite{Mandel-91}, in a paper which introduces the quantitative concept of degrees of indistinguishability. More than that, Mandel's degrees of indistinguishability are identical to degrees of coherence in a precise framework. That means a comprehensive understanding of the double-slit experiment can bring us some light about the wave/particle duality of photons in a way which allows us to equal degrees of coherence to degrees of path indistinguishability. So, my first concern here refers to those situations where indistinguishability plays an important role in quantum theories. Next I analyze those situations from a quasi-set-theoretic point of view. Why is that? Because quasi-set theory was conceived from a philosophical perspective which was committed to Heinz Post's ideas. And Post's ideas were strongly motivated by those problems regarding indistinguishability in quantum mechanics.

In quasi-set theory $\cal Q$ there is a primitive binary relationship $\equiv$ called {\em indistinguishability\/} and a defined binary relationship $=_E$ called {\em extensional identity\/}. Besides, terms in $\cal Q$ are either collections or atoms. Atoms are empty terms which are not collections. So, the philosophical content here is as follows. If $x\equiv y$ we say there is no way to distinguish $x$ and $y$. If $x$ and $y$ are atoms such that $x\equiv y$ entails $x=_E y$, we say both $x$ and $y$ are the very same individual. And such individuality is provided by extensional identity (as it happens in ZFU, where individuality is granted through identity). If $x\equiv y$ does not entail $x=_E y$, we say that although $x$ is indistinguishable of $y$, that fact does not allow us to say that either $x$ or $y$ is an individual. In that case we are talking about non-individuals, in a very precise way provided by the axioms of $\cal Q$. In \cite{Benda-18} and \cite{Sant'Anna-19} there are some philosophical discussions about the explanatory power of quasi-sets. Nevertheless, in this paper I am focused on the relations between quasi-sets and quantum physics. Some authors \cite{Muller-08} advocate that non-standard approaches for dealing with the problem of (non)individuality (like quasi-set theory and alike) are unnecessary. But necessary or not, the fact is that there is a literature about this subject, which cannot be ignored. And in this paper I follow a different approach for discussing Krause's proposal. Even if we assume a quasi-set-theoretic approach for dealing with identity issues, such a formal framework is limited to very specific cases of quantum phenomena. In the next sections I discuss such a limitation.

In other words, in this paper I am raising the following question: {\em Does quasi-set theory provide a comprehensive description into how indistinguishability is related to identity within non-relativistic quantum physics?\/}

Since the early days of quantum studies it is well known that coherence, in interference experiments, is somehow related to the indistinguishability of particle trajectories. If a photon is detected by a photodetector, such a particle can come from either one of the two secondary sources in a double-slit experiment. And if those two possible paths are indistinguishable, then the probability amplitude for the photon to be detected is the sum of the probability amplitudes associated with the two possible paths. The detection probability, which is the square modulus of the probability amplitude, then exhibits interference - a well known phenomenon. Since there may be a precise definition for degrees of coherence, why can't we think about {\em degrees of indistinguishability\/}? And if that is the case, how can quasi-set theory help us to cope with such degrees of indistinguishability?

The quasi-set theoretic formalism for Fock Spaces \cite{Domenech-08} \cite{Domenech-10} is relevant in our discussion in the next sections. After all, within Fock Spaces it is possible to define creation and annihilation operators. And I show in this paper that it is exactly within this field-theoretic framework (Fock Spaces) that quasi-set theory fails to relate indistinguishability (between certain quantum objects) with identity (of quantum objects).

From a sociological point of view, there are at least two possible methodologies for formally dealing with the relationships between indistinguishability among quantum objects and their individuality. The first approach starts with a metaphysical assumption which grounds all the necessary intuitions for the development of at least one formal system. And that formal system is somehow expected to provide a mapping between theoretical concepts and experiments. The second approach starts with a catalogue of all available and relevant experimental results. Those experimental data provide all necessary intuitions for the development of at least one formal system which, again, is supposed to provide a mapping between theoretical concepts and experiments. Once the theory is developed, a metaphysical interpretation is almost unavoidable. I recognize there is a vague distinction here concerning those methodologies. After all, any metaphysical assumption for a physical theory is supposed to be based on experimental data. But clearly quasi-set theory was developed according to the first one, in the sense that some remarkable experimental data are simply ignored, for example, in \cite{French-06}. Some of those data refer to Mandel's degrees of indistinguishability \cite{Mandel-91} in quantum optics and violations of the Symmetrization Postulate \cite{Angelis-96} \cite{Modugno-98}. That is why the quasi-set-theoretic formalism for Fock spaces is relevant in our discussion. Papers like \cite{Domenech-08} and \cite{Domenech-10} may suggest to their readers that quasi-set theory is consistent with modern views regarding quantum theories. So, while authors like Muller and Saunders \cite{Muller-08} consider non-standard formal systems for indistinguishability in quantum mechanics are unnecessary, in this paper I simply show that at least one of those non-standard formal systems (namely, $\cal Q$) does not provide a comprehensive picture for quantum mechanics. That means this is a paper on the epistemological character of quasi-sets.

At the end of the paper I briefly discuss a possible way to cope with indistinguishability issues without abandoning identity (in a sense). And the idea suggested here is quite different from the one proposed by Muller and Saunders \cite{Muller-08}. In \cite{Muller-08} the authors are concerned with Leibniz’s Principle of the Identity of Indiscernibles. They discern fermions by means of relations which are independent of quantum-mechanical probabilities. Nevertheless, in principle, there is at least one way to formally describe indistinguishability among quantum objects without any explicit reference to properties of those objects (whether relational or not). And that can be achieved by focusing on the underlying logic of the theory which is supposed to bring some light about indistinguishability.

So, my goal here can be briefly presented as it follows. The double-slit experiment shows quasi-set theory does not provide a comprehensive picture for the relationship between indistinguishability and identity in the quantum realm. That happens because the double-slit experiment reveals a close relationship between indistinguishability of photon paths and coherence of fields. And quasi-set theory does not allow us to take into account the wave/particle duality and its role on the understanding of indistinguishability, particularly if we intend to address the concepts of individuality versus non-individuality. Quasi-set theory is somehow focused on a limited view regarding quantum objects. That happens due to a strong philosophical commitment concerning the concept of identity. But it is possible that our attention should be addressed to other logical issues rather than identity, as I finally discuss at the end of the paper.

\section{Wave/Particle Duality}

French and Krause claim that ``standard set theories do not provide adequate mathematical tools for dealing with collections of `legitimate' indistinguishable entities'' \cite{French-06}. In order to support their claim they appeal to some sort of comparison between Schr\"odinger's ideas (concerning the quanta) and the usual ways identity is formalized in standard mathematics. According to Schr\"odinger, ``It is not at all easy to realize this lack of individuality and to find words for it''. And standard set theories, on the other hand, are strongly grounded on a rigid notion of identity. So, another kind of logical apparatus is provided through quasi-set theory.

However, in 1991 - many years after Heinz Post published his famous paper on (non)individuality \cite{Post-63} - Leonard Mandel published an intriguing and insightful theoretical approach regarding the relationship between indistinguishability (of secondary sources in the double-slit experiment) and coherence \cite{Mandel-91}. He proved there may be a mathematical identity between indistinguishability and coherence (which was inspired on experimental results \cite{Zou-91}). And since coherence admits degrees in a very precise sense, the same takes place with indistinguishability. That result suggests the problem of non individuality (in the sense that individuality is related to indistinguishability) is a little more sophisticated than just considering collections of indiscernible objects (as Krause and French discuss in their book). {\em The concept of indistinguishability (which seems to be relevant for quantum trajectories, besides particles) seems to be directly related to the particle/wave duality in quantum phenomena\/}. And if that is true, quasi-set theory does not provide a complete explanatory picture for the problem of non-individuality in quantum physics.

The reader could consider that my concern (stated in the previous paragraph) is not legitimate due to the fact that quantum trajectories cannot be considered quantum objects (in some way). Nevertheless, quantum trajectories in the double slit experiment refer to field sources, whether they are secondary or not. So, it makes sense to assume those trajectories are somehow related to objects, namely, field sources.

Before introducing Mandel's theoretical proposal, let me present his motivations, which were based on a rather peculiar experiment \cite{Zou-91}. Interference experiments with single photons are commonly performed by means of a technique called {\em parametric down-conversion\/}. The general idea is something like this. Nonlinear crystals (the parametric down-converters) are pumped by a high energy photon which is converted into a pair of lower energy photons, namely, the signal photon and the idler photon (principles of energy conservation and momentum conservation are taken into consideration here). That is a standard procedure for detection of second order interference (i.e., to study correlations between field intensities between two points) of signal photons. And such a technique is largely used for the generation of entangled photon pairs. And Mandel's mathematical results in \cite{Mandel-91} were strongly motivated by experimental results achieved with a specific experimental setup involving two coherently parametric down-converters which were pumped by a single photon. For details see \cite{Zou-91}. So, here follows a brief description of the experiment.

In \cite{Zou-91}, the authors work with two similar nonlinear crystals in a Mach-Zehnder-style interferometer. A single photon from an Argon laser is sent to a beam splitter, which allows two possible paths. One path goes to one nonlinear crystal and the other one goes to the other crystal. Those crystals (1 and 2) decompose the incident photon into a pair of a signal photon and an idler photon (for each crystal). Therefore, we have two possible signal photon paths and two possible idler photon paths. The main goal of the experiment was to analyse interference associated to the detected signal photons (whose trajectories come together at a second beam splitter), as long the trajectories of the idler photons are aligned. And Mandel and collaborators found the following results: (i) The signal photons are coherent (interfere) even if there is no induced emission from the second crystal (remember signal photons and idler photons are supposed to be in an entangled state); and (ii) if the connection between the idler photons is broken (by either misalignment or an opaque obstacle between the crystals), then we have a correlation between signal photon paths and coherence of signal photons. That correlation is as follows. When the connection between idler photons is broken, it is possible to determine if the detected signal photon came from crystal 1 or crystal 2, by counting coincidences of detection between the signal detector and the idler detector. If both detectors register photons, then we know the signal photon came from crystal 2. If the signal detector registers a photon but the idler detector does not, then we know the signal photon came from crystal 1. For details, as already said, see \cite{Zou-91}. And this possibility of distinguishing paths wipes out interference in the signal detector.

Mandel and collaborators interpreted their results as a consequence of the relationship between indistinguishability of the signal photon paths and coherence. So, what is Mandel's theoretical proposal?

By considering the simplest case of interference of two single-mode fields from two secondary sources, and the detection of a single photon, Mandel starts with the quantum state of a light beam given by

$$|\psi\rangle = \alpha |1\rangle_1|0\rangle_2 +\beta |0\rangle_1|1\rangle_2,$$

\noindent
where $|n\rangle _i$ stands for the quantum state in which $n$ photons originate from secondary source $s_i$, and, of course,

$$|\alpha|^2 + |\beta|^2 = 1.$$

In that case we have an entangled quantum state in which the photon can originate either in source $s_1$ with probability $|\alpha|^2$ or in source $s_2$ with probability $|\beta|^2$. Nevertheless, those two possibilities may be simply indistinguishable, if we are talking about a coherent mixture of states. What do we mean by that? From an intuitive point of view, it means the detection of a single photon in the photodetector does not provide enough information regarding that photon's path through the double-slit. From a deeper point of view, those possibilities are strongly related to the coherence between both secondary sources, as I discuss below.

Since density matrices are remarkably helpful for describing both pure and mixed states, Mandel makes use of the density operator for the double-slit experiment. In the case of an incoherent mixture we have a simple diagonal operator

$$\hat\rho_D = |\alpha|^2 |1\rangle_1|0\rangle_{2\;2}\langle 0|_1\langle 1| + |\beta|^2 |0\rangle_1|1\rangle_{2\;2}\langle 1|_1\langle 0|,$$

\noindent
where $D$ stands for {\em distinguishability\/}. Coefficients $|\alpha|^2$ and $|\beta|^2$ are still probabilities that the photon comes from secondary source $s_1$ or secondary source $s_2$, respectively. But since we are talking about an incoherent mixture, such probabilities may be (in principle) distinguished. As remarked by Mandel, ``in principle, there exists some experimental scheme that allows the source of a detected photon to be identified.'' Therefore, the problem of distinguishing trajectories is now just an experimental challenge (like that one analysed in \cite{Zou-91}).

In the case of a coherent mixture, we have another density operator, this time non-diagonal. Its Fock expansion is given by:

$$\hat\rho_{ID} = |\alpha|^2 |1\rangle_1|0\rangle_{2\;2}\langle 0|_1\langle 1| + |\beta|^2 |0\rangle_1|1\rangle_{2\;2}\langle 1|_1\langle 0| +$$
$$\alpha\beta^*|1\rangle_1|0\rangle_{2\;2}\langle 1|_1\langle 0| + \mbox{h.c.},$$

\noindent
where h.c. stands for the Hermitean conjugate of the previous term and $ID$ is an abbreviation for {\em indistinguishability\/}.

So, a quantum system prepared in $\hat\rho_{ID}$ (photon paths are indistinguishable) is capable of showing interference, while a system prepared in $\hat\rho_D$ (photon paths are distinguishable) is not.

Next Mandel considers an arbitrary one-photon state with density operator

$$\hat\rho = \rho_{11}|1\rangle_1|0\rangle_{2\;2}\langle 0|_1\langle 1| + \rho_{22}|0\rangle_1|1\rangle_{2\;2}\langle 1|_1\langle 0| +$$
\begin{equation}
\rho_{12}|1\rangle_1|0\rangle_{2\;2}\langle 1|_1\langle 0| + \mbox{h.c.}\label{firstmatrices}
\end{equation}

This last operator $\hat\rho$ can be decomposed in a unique form as it follows:

\begin{equation}
\hat\rho = P_{ID}\hat\rho_{ID} + P_D\hat\rho_D,\label{matrices}
\end{equation}

\noindent
where

\begin{equation}
P_{ID} + P_D = 1.\label{PIDvariando}
\end{equation}

$P_{ID}$ and $P_D$ stand, respectively, for the probabilities for the sources to be intrinsically indistinguishable and distinguishable. Mandel presents a geometrical interpretation for such a result in terms of Bloch vectors. For details, see his paper.

Equation \ref{matrices} is an identity between matrices. So,
$$\rho_{11} = |\alpha|^2,$$
$$\rho_{22} = |\beta|^2,$$
$$\rho_{12} = P_{ID}\alpha\beta^*.$$

That means we can rewrite both $P_{ID}$ and $\alpha\beta^*$ solely in terms of matrix elements as it follows:

$$\alpha\beta^* = \sqrt{\rho_{11}\rho_{22}}\exp(i\,\mbox{arg}\,\rho_{12}),$$

$$P_{ID} = \frac{\rho_{12}}{\sqrt{\rho_{11}\rho_{22}}}\exp(-i\,\mbox{arg}\,\rho_{12}) = \frac{|\rho_{12}|}{\sqrt{\rho_{11}\rho_{22}}}.$$

$P_{ID}$ is the degree to which the two sources are intrinsically indistinguishable in the general quantum state $\hat\rho$ given by Equation \ref{firstmatrices}. Thus, $P_{ID} \in [0,1]$.

An alternative theoretical solution for the problem of the physical interpretation of $P_{ID}$ may be found, e.g., in an independent work by Rabinowitz in \cite{Rabinowitz-95}. This author takes into account the particle's momentum components in order to allow identification of the secondary source. From an experimental point of view, this idea is somehow a good one. In \cite{Koksis-11}, for example, the authors sent single photons emitted by a quantum dot through a double-slit interferometer and reconstructed these trajectories by performing a weak measurement of the photon momentum. But in the case of Mandel's proposal there is no need to consider momentum \cite{Zou-91}.

In order to relate $P_{ID}$ to the coherence properties of the source field, Mandel uses a Fourier decomposition of the total field operator $\hat{E}(r_j)$ -- where $j = 1,2$ -- into its positive- and negative-parts $\hat{E}^{(+)}(r_j)$ and $\hat{E}^{(-)}(r_j)$, respectively. By assuming that

$$\hat{E}^{(+)}(r_j) = K\hat{a},$$

\noindent
where $K$ is a complex constant and $\hat{a}$ is a photon annihilation operator, now it is possible to use Equation \ref{firstmatrices} in order to get the (second order) mutual coherence function $\Gamma_{12}^{(1,1)}$ for the secondary sources:

$$\Gamma_{12}^{(1,1)} = \langle \hat{E}^{(-)}(r_1)\hat{E}^{(+)}(r_2)\rangle = |K|^2\mbox{Tr}(\hat{a}_1^\dagger\hat{a}_2\hat\rho) = |K|^2\rho_{21},$$

\noindent
where $\langle \cdots \rangle$ stands for statistical average and $\dagger$ denotes Hermitean conjugate. Besides,

$$\Gamma_{11}^{(1,1)} = |K|^2\rho_{11} = |K|^2|\alpha|^2$$

\noindent
and

$$\Gamma_{22}^{(1,1)} = |K|^2\rho_{22} = |K|^2|\beta|^2$$

Finally, the normalized mutual coherence function is given by

$$\gamma_{12}^{(1,1)} = \frac{\Gamma_{12}^{(1,1)}}{\Gamma_{11}^{(1,1)}\Gamma_{22}^{(1,1)}} = \frac{\rho_{21}}{\sqrt{\rho_{11}\rho_{22}}}.$$

In other words,

\begin{equation}
|\gamma_{12}^{(1,1)}| = P_{ID}.\label{greatresult}
\end{equation}

This last equation is a major result obtained by Mandel, since it quantifies (for the first time) a very old belief among physicists. It says the degree of coherence is also the degree of intrinsic indistinguishability of the two secondary sources.

Actually, in his paper Mandel goes further. He shows the degree of intrinsic indistinguishability $P_{ID}$ is also related to the fringe visibility $\cal V$ in an interference experiment ($P_{ID} \leq {\cal V}$). For details see \cite{Mandel-91} and \cite{Zela-14}.

In \cite{Mandel-95} Mandel improved his results by considering a two-photon interference. And once again it was confirmed an equality between a degree of coherence and a degree of indistinguishability. So, in principle, Mandel's results can be extended to an arbitrary number of photons, despite the fact that any mathematical modelling of such a many-particle-system is in no way an easy task to be performed.

An intriguing consequence from Mandel's mathematical account is the fact that it suggests something which may be going beyond experimental results. The experimental setup in \cite{Zou-91} works almost like an on/off switch, in the sense of a broken/not-broken connection between idler photons' paths. Nevertheless, we cannot forget we can manage to deal with slight misalignments between idler photons' paths. Besides, it makes perfect sense talking about degrees of fringes visibility in the double-slit experiment. And in the sense given above, a notion of degrees of indistinguishability (equal to a degree of coherence) is consistent with usual formalisms for non-relativistic quantum field theory. More recently Coles and collaborators \cite{Coles-14} demonstrated that wave-particle duality principles precisely correspond to a modern formulation of the Uncertainty Principle. That means the notion of partial indistinguishability between photon paths (in interferometers) is a quite common interpretation nowadays. For details, see \cite{Coles-14} and its references.

\section{Quasi-Sets}

In this Section I briefly discuss the Symmetrization Postulate within Quasi-Set Theory $\cal Q$. Besides, this brief review on $\cal Q$ is necessary for the discussion about Mandel's degrees of indistinguishability in the next Section.

Quasi-set theory $\cal Q$  \cite{Krause-05} is a first order theory without identity \cite{Mendelson-97}, strongly committed to ZFU (Zermelo-Fraenkel set theory with {\em Urelemente\/}) like axioms. Its primitive concepts are three unary predicate letters $m$, $M$, and $Z$, two binary predicate letters $\in$ and $\equiv$, and one unary function letter $qc$. Intuitively speaking, all terms of $\cal Q$ are either collections (qsets) or atoms ({\em Urelemente\/}). If $x$ is an atom, then $x$ is called either a micro-atom ($m(x)$) or a macro-atom ($M(x)$). And the axioms of $\cal Q$ forbid the existence of any term which is a micro-atom and a macro-atom. If $Z(x)$, then $x$ is a specific quasi-set which corresponds to a standard set of ZFU, in a very precise sense. If $x\in y$, we say $x$ belongs to $y$. We can say that $x$ is an element of $y$ as well. If $x\equiv y$, we say $x$ and $y$ are indistinguishable terms, or simply indistinguishable. Finally, $qc(x)$ is referred to as the quasi-cardinality of $x$. The intuitive idea concerning $qc$ is that of ``number of elements of $x$'' (whether it is finite or transfinite). In other words, even in a formal theory without identity it is still possible to give a precise sense about quantities.

The next definitions in $\cal Q$ give us a first formal glimpse about quasi-sets. The symbol $\vdots$ stands for a metalinguistic symbol used here to separate the {\em definiendum\/} and the {\em definiens\/} in each definition.

\begin{definicao}

$Q(x) \; \vdots \; \neg (m(x) \vee M(x))$

\end{definicao}

We read $Q(x)$ as ``$x$ is a quasi set''. In other words, no quasi-set is an atom.

\begin{definicao}

$x =_E y \; \vdots \; (Q(x) \wedge Q(y) \wedge \forall z (z\in x \Leftrightarrow z\in y)) \vee (M(x) \wedge M(y) \wedge \forall_Q z (x\in z \Leftrightarrow y\in z)$\label{definicao2}

\end{definicao}

$\forall_Q z$ is the universal quantifier bounded to the predicate letter $Q$. This last definition refers to {\em extensional identity\/} $=_E$. The intuitive idea concerning the formula $x =_E y$ is as it follows. Term occurrences $x$ and $y$ are extensionally identical iff they are either (i) collections sharing the same elements or (ii) macro-atoms belonging to the same collection.

The first axioms of $\cal Q$ are the following:

\begin{itemize}

\item[\bf Q1] $\forall x (x \equiv x)$

\item[\bf Q2] $\forall x \forall y (x \equiv y \Rightarrow y \equiv x)$

\item[\bf Q3] $\forall x \forall y \forall z ((x \equiv y \wedge y \equiv z) \Rightarrow x \equiv z)$

\item[\bf Q4] $\forall x \forall y (x =_E y \Rightarrow (A(x,x) \Rightarrow A(x,y)))$, where $A(x,x)$ is a formula in $\cal Q$ and $A(x,y)$ is obtained from $A(x,x)$ by replacing at least one of the free occurrences of $x$ by $y$, if $y$ is free for $x$ in $A(x,x)$.

\end{itemize}

The first three axioms state indistinguishability is an equivalence relation. {\bf Q4}, on the other hand, is an axiom schema (for each formula $A$, as long as it satisfies the conditions given above, there is a corresponding axiom following {\bf Q4}) which may be considered as the heart and soul of quasi-set theory.

In ZFU we have very similar axioms. Identity $=$, in that theory, is an equivalence relation as well. But there is an axiom schema in ZFU which states the following:

\begin{equation}
\forall x \forall y (x = y \Rightarrow (A(x,x) \Rightarrow A(x,y))),\label{substituicao}
\end{equation}

\noindent
where $A(x,x)$ is a formula in ZFU and $A(x,y)$ is obtained from $A(x,x)$ by replacing at least one of the free occurrences of $x$ by $y$, if $y$ is free for $x$ in $A(x,x)$.

Such a scheme is known as the {\em substitutivity axiom\/}. Notwithstanding, {\bf Q4} is some sort of restriction on the use of $\equiv$. Substitutivity $(A(x,x) \Rightarrow A(x,y))$, in $\cal Q$, can take place only among those very specific cases where $x \equiv y$ entails $x =_E y$. Within quasi-set theory $x =_E y$ necessarily entails $x \equiv y$. But the converse is not always valid since, according to definition \ref{definicao2}, $x =_E y$ demands very specific conditions for $x$ and $y$. In other words, by comparing {\bf Q4} with equation \ref{substituicao}, it is clear that extensional identity $=_E$ in $\cal Q$ has all the features of standard identity $=$ in ZFU. And that is a key point. Quasi-set theory was philosophically inspired on Post's ideas. Thus, extensional identity is not a limit case of indistinguishability. What does that mean? It means the universe of discourse in $\cal Q$ clearly separates terms into two disjoint classes (I am using the term ``class'' in an intuitive sense), namely, those terms who behave like terms in ZFU (terms $x$ such that either $M(x)$ or $Z(x)$) and those who do not copy ZFU. This last class corresponds to the non-standard counterpart of $\cal Q$. In other words, there is no continuum transition from indistinguishability to extensional identity. There is no continuum transition from the non-standard counterpart of $\cal Q$ to its classical counterpart.

In \cite{Krause-05}, e.g., the authors prove the following result: if axiom scheme {\bf Q4} is replaced by a weaker statement which asserts that substitutivity $(A(x,x) \Rightarrow A(x,y))$ can take place only when $x \equiv y$ and neither $x$ nor $y$ is a micro-atom, then this new version of quasi-set theory is equivalent to ZFU. That means indistinguishability is simply reduced to standard identity if we promote such a modification on {\bf Q4}. In other words, substitutivity $(A(x,x) \Rightarrow A(x,y))$ is a very sensitive issue in the quasi-set-theoretic framework, when $x \equiv y$, in the sense there is no continuum transition from indistinguishability to extensional identity. Depending on the way we state axiom {\bf Q4} we may collapse $\cal Q$ to standard ZFU, and micro-atoms behave exactly like macro-atoms. That fact confirms my claim that extensional identity is not a (continuum) limit case for indistinguishability. Any modification in {\bf Q4} demands a discrete transition (by changing its statement) from one old version into a new one.

All those features concerning $\cal Q$ confirm that quasi-set theory was quite successful as a formal translation of one of the main ideas of Heinz Post, namely, that non-individuality of quantum particles should be assumed {\em right at the start\/}.

Permutations are not observable, according to the next theorem in $\cal Q$. For understanding its statement, it is important to know that $[z]$ corresponds to the qset of all terms which are indistinguishable from $z$; and $z'$ is a qset whose quasi-cardinality is extensionally identical to $1$ and such that $z'\subseteq [z]$. Besides, in $\cal Q$ there is an axiom for union which allows us to write $x\cup y$ for the union between qsets $x$ and $y$.

\begin{teorema}
Let $x$ be a finite qset (its quasi-cardinality is a natural number) such that $x\neq_E [z]$ and $z$ is an $m$-atom such that $z\in x$. If $w\equiv z$ and $w\not\in x$, then there exists $w'$ such that $(x-z')\cup w' \equiv x$.
\end{teorema}

In order to prove this theorem, the authors in \cite{French-06} consider two scenarios: first, the case where $t\in z'$ does not belong to $x$; second, the case where $t\in z'$ does belong to $x$. Then they conclude that, for $x$ finite, any `exchange' (observe the quotation marks, which are originally used by French and Krause) of elements $z$ by the corresponding indistinguishable elements $w$ results a qset which is indistinguishable from $x$. The problem here, however, is the fact that $\cal Q$ cannot decide if there was indeed any permutation at all. In the first scenario, the proof seems to suggest there was no permutation. In the second one we are induced to believe there was a permutation. But both of them produce the same result, in the sense that in both situations we have $(x-z')\cup w'$ is indistinguishable from $x$. So, what is the difference between (i) a metaphysical view in which permutations do take place but are not observable and (ii) a metaphysical view in which permutations simply do not occur? If the Symmetrization Postulate is not clear enough - when expressed in intuitive terms - quasi-set theory does not improve that situation. Invariance principles in physics are always verifiable. But that does not happen here, within $\cal Q$. Since there is physical evidence of violation of SP \cite{Angelis-96} \cite{Modugno-98}, quasi-set theory seems to offer an undesirable solution for its formal statement. After all, are there degrees of indistinguishability which could explain those slight violations of the SP? A useful analogy can be made with the concept of force in classical mechanics. Newton's second law refers, in principle, to resultant forces and individual forces. But how can we make the distinction between a resultant force and an individual one? That's why Hertz \cite{Hertz-56} proposed a new mechanics, in the 19th century, where the notion of force is not among its primitive concepts \cite{Sant'Anna-96} \cite{Sant'Anna-03}. That happened because Hertz understood the notion of force has a strong metaphysical commitment which is hardly justifiable in physical terms. Since the permutation between individual forces and resultant forces are not observable, how can we guarantee that any inertial body is not being pushed by angels in such a way that the resultant force from those angels is null?

In \cite{Domenech-10} the authors show how to avoid the label-tensor-product-vector-space-formalism of quantum field theory when dealing with indistinguishable quanta, by using quasi-set theory. According to them, ``states in this new vector space, that we call the Q-space, refer only to occupation numbers and permutation operators act as the identity operator on them, reflecting in the formalism the unobservability of permutations, a goal of quasi-set theory''. Nevertheless, in that paper (and in its preceding work in \cite{Domenech-08}) the authors ignore the fact that indistinguishability in quantum theories may refer to much more than just the unobservability of permutations. So, their quasi-set-theoretic approach to Fock spaces does not necessarily take into account all quantum phenomena where indistinguishability plays a significant role. Consequently, that fact may seriously undermine any ambition for the understanding of the problem of non-individuality in the quantum realm, as I discuss in the next Section.

\section{Which-Way Information and Quasi-Sets}

This section presents the first logical and philosophical difficulties of quasi-set theory to cope with Mandel's degrees of indistinguishability. In the next Section I discuss this subject in a more detailed way, by suggesting there is no need to postulate that indistinguishability entails any sort of non-individuality.

According to Mario Rabinowitz \cite{Rabinowitz-95}, ``[t]he wave-particle duality is the main point of demarcation between quantum and classical physics, and is the quintessential mystery of quantum mechanics.'' Such a statement may be easily seen as an exaggeration. After all, quantum theories in general present many unexpected features if we compare them to classical physics. The problem of non-individuality itself is obviously one of them. But we could recall other aspects, like the inherent probabilistic character of quantum phenomena, non-locality, non-realism, interaction-free measurements, and so on. Nevertheless, there is no doubt the wave-particle duality plays a major role on the distinction between quantum mechanics and classical mechanics. Mandel's work (presented above) is made within a field-theoretic framework, since he employs the annihilation operator $\hat a$. Nevertheless, the annihilation operator is used just in order to get interference fringes in a non-relativistic setting. So, {\em mutatis mutandis\/}, this non-relativistic quantum field theory is consistent with standard quantum mechanics.

According to French and Krause \cite{French-06}, a considerable body of evidence from quantum mechanics points to a rather peculiar perspective regarding individuality: quantum particles are somehow devoid of individuality, while classical particles are endowed with some sort of ``transcendental individuality'' (if we use Post's terminology). That philosophical view was inspired on key ideas by Heinz Post, Erwin Schr\"odinger, Yuri Manin, and many others. And that view has a metaphysical appeal \cite{Manin-76}. So, quasi-set theory was introduced as a genuine formal framework for dealing with this metaphysical perspective. That is why the notion of indistinguishability $\equiv$ was introduced in the sense of axioms {\bf Q1-Q4} (and other axioms) of theory $\cal Q$.

Thus, let me start with equation \ref{PIDvariando}. Is it possible to provide a quasi-set-theoretic perspective for $P_{ID}$ in equation \ref{PIDvariando}? My answer is positive, if we are looking for a plain and simple quasi-set-theoretical picture for Mandel's equations (like equation \ref{greatresult}). After all, in \cite{Domenech-08} and \cite{Domenech-10} the authors show how to rewrite Fock space formalism within the non-standard counterpart of quasi-set theory. That happens because all they wanted was to dismiss the Symmetrization Postulate from quantum theory formalism. On the other hand, my answer is negative if we are looking for explanatory power in quasi-set theory, even from an ontological perspective. And that was, since its birth, the main goal proposed by Krause and collaborators: to provide explanatory power for the problem of non-individuality in quantum theories. So, my answer is negative simply because the problem of non-individuality seems to be related to something else besides the Symmetrization Postulate.

According to equation \ref{PIDvariando},

$$P_{ID} + P_D = 1,$$

\noindent
where both terms take values in the $[0,1]$ real interval. So, if $s_1$ and $s_2$ denote the secondary sources, we could translate this (within the language of $\cal Q$) as

\begin{equation}
s_1 \equiv s_2\;\mbox{iff}\; P_{ID} = 1,\label{baca}
\end{equation}

\noindent
which is equivalent to say

$$s_1 \equiv s_2\;\mbox{iff}\; P_{D} = 0.$$

But what about all other possible real values for $P_{ID}$? The logical axioms of $\cal Q$ are the same of classical logic. So, all we have is

$$\neg(s_1 \equiv s_2)\;\mbox{iff}\; P_{ID} \neq 1.$$

In other words, all descriptive power of equation \ref{PIDvariando} is just missing in $\cal Q$. If $P_{ID} = .5$, then all we know from $\cal Q$ is that $\neg(s_1 \equiv s_2)$. The same happens for all other possible values for $P_{ID}$, as long $P_{ID}$ is not 1. So, what are the proposed quasi-set-theoretical ontological consequences for cases like $P_{ID} \neq 1$? How is this related to non-individuality?

In order to avoid that inconvenience, we could recall that quasi-set theory $\cal Q$ has a copy of ZFU within its universe of discourse. So, equation \ref{PIDvariando} can certainly be expressed in $\cal Q$, if we employ this ZFU counterpart of $\cal Q$. Nevertheless, a copy of ZFU within $\cal Q$ is just standard mathematics. So, what about the non-standard part of $\cal Q$? If quasi-set theory is supposed to provide the mathematics of non-individuality (as stated by French and Krause in chapter 7 of their book \cite{French-06}), such a mathematical framework should take into account those degrees of indistinguishability which naturally occur in the double-slit experiment.

Another possible strategy to deal with equation \ref{PIDvariando} in $\cal Q$ is by assuming

$$s_1 \equiv s_2\;\mbox{iff}\; P_{ID} \neq 0.$$

But once again we loose the descriptive power of equation \ref{PIDvariando}.

What do I mean by ``descriptive power''? Mandel's results suggest something like a continuum spectrum of ``levels of indistinguishability''. And those levels of indistinguishability are measurable by means of fringes visibility in the double-slit experiment. But according to axiom {\bf Q4} there is no such a thing like ``levels of indistinguishability'' in $\cal Q$. That happens because two arbitrary terms $x$ and $y$ are either indistinguishable ($x \equiv y$) or distinguishable ($\neg (x\equiv y)$). And that is all! Besides, extensional identity $=_E$ in $\cal Q$ (which is equivalent to identity $=$ in ZFU) is no continuous limit at all of indistinguishability $\equiv$, as already discussed.

Notwithstanding, for the sake of argument let me dismiss for a while the concept of extensional identity. After all, if $x$ and $y$ are distinguishable (i.e., $\neg(x\equiv y)$), that fact does not entail $x =_E x$ (or $y =_E y$). In other words, even when we are talking about distinguishable terms of $\cal Q$, that does not necessarily mean those terms are provided by some sort of identity. Therefore, in order to accommodate Mandel's results in $\cal Q$ there is no need to assume extensional identity is supposed to be a continuum limit of indistinguishability. So, what happens if we focus our attention on indistinguishability alone?

Since quasi-set theory allows us to define a natural extension of the concept of function in ZFU \cite{Krause-92} \cite{Krause-05}, what about a quasi-set-theoretic notion of {\em fuzzy\/} indistinguishability? If that is possible, then there is a good chance of talking about levels of indistinguishability, as required by Mandel's results (which, by the way, were extended even for light polarization \cite{Zela-14}).

Roughly speaking, a quasi-function with domain $x$ and codomain $y$ is a quasi-set $f$ of ordered pairs $\langle a, b\rangle$ such that: (i) for all $a\in x$ there is an image $b\in y$ such that $\langle a, b\rangle \in f$; and (ii) if $\langle a, b\rangle \in f$, $\langle a', b'\rangle \in f$, and $a\equiv a'$, then $b\equiv b'$. In other words, any pair of indistinguishable elements from domain $x$ have indistinguishable images in the codomain $y$.

It is worth recalling again there is a copy of ZFU in $\cal Q$. Therefore, it is possible to define standard real numbers $\Re$ in $\cal Q$, where identity $=$ in $\Re$ is replaced by extensional identity $=_E$. More than that, standard inequality relations between real numbers like $<$, $>$, $\leq$, $\geq$, and $\neq$ are replaced, respectively, by their quasi-set-theoretic counterparts $<_E$, $>_E$, $\leq_E$, $\geq_E$, and $\neq_E$. Within the ZFU-copy of real numbers $\Re$ in $\cal Q$, I am using the standard notation for real numbers operations like sum ($+$), difference ($-$), and multiplication.

I hope the reader realizes I am not loosing my focus on indistinguishability, as proposed above. I am just trying to use the ZFU counterpart of $\cal Q$ in order to allow a definition of degrees of indistinguishability. But I am not trying to get extensional identity as some sort of continuum limit of indistinguishability, since it was already argued that such a task is not doable.

Thus, one possible quasi-set-theoretic solution for the notion of ``degree of indistinguishability'' is accomplished by the following definitions.

\begin{definicao}

$\langle M, d\rangle$ is a {\em quasi-metric space\/} iff

\begin{itemize}

\item[QM1)] $M$ is a non-empty quasi-set.

\item[QM2)] $d:M\times M\to \Re$ is a quasi-function such that for each ordered pair $\langle a,b\rangle$, there is an image $d\langle a,b\rangle$ called the {\em quasi-distance\/} between $a$ and $b$.

\item[QM3)] $d\langle a,b\rangle \geq_E 0$.

\item[QM4)] $d\langle a,b\rangle =_E 0$ iff $a\equiv b$.

\item[QM5)] $d\langle a,b\rangle =_E d\langle b,a\rangle$.

\item[QM6)] $d\langle a,c\rangle \leq_E d\langle a,b\rangle + d\langle b,c\rangle$

\end{itemize}

\end{definicao}

This last definition is a natural generalization of metric spaces in ZFU. If $M$ is a ZFU-set (i.e., $Z(M)$), then all occurrences of $\equiv$ may be replaced by $=_E$; and so we have a standard definition of metric space.

Now consider a special case of quasi-metric space such that the codomain of $d$ is the interval $[0,1]\subset \Re$. We can call this a {\em differentiation quasi-metric space\/}. So,

\begin{definicao}\label{graudeindistinguibilidade}

If $\langle M, d\rangle$ is a differentiation quasi-metric space, and $a\in M$ and $b\in M$, then

$$a\equiv_r b \Leftrightarrow r =_E 1 - d\langle a, b\rangle$$

\end{definicao}

A straightforward consequence of definition \ref{graudeindistinguibilidade} is the theorem below:

\begin{equation}
a\equiv_1 b \Leftrightarrow a\equiv b.\label{beca}
\end{equation}

That is easily proven from axiom $QM4$. Besides, if $r \neq_E 1$, then

\begin{equation}
a\equiv_r b \Leftrightarrow \neg(a\equiv b).\label{cerne}
\end{equation}

So, thanks to this last definition, I have defined a whole class (the term ``class'' here is used in a merely intuitive sense, since there are no classes in $\cal Q$ and binary relations are no terms) of binary relations $\equiv_r$. In other words, the arbitrary one-photon state given by equation \ref{firstmatrices} could be somehow ingeniously associated to a quasi-set $M$ of possible secondary sources such that

$$s_1 \equiv_r s_2 \Leftrightarrow P_{ID} =_E r,$$

\noindent
which is consistent with equations \ref{baca} and \ref{beca}.

But what can the axioms of $\cal Q$ elucidate about $\equiv_r$ when $r\neq_E 1$? Unfortunately, nothing at all. And equation \ref{cerne} proves my claim. If $r\neq_E 1$, then $a\equiv_r b$ is equivalent to $\neg(a\equiv b)$ within the context of a differentiation quasi-metric space. And that's it!

From a purely syntactic point of view, the axioms of $\cal Q$ do not provide any insight about $\equiv_r$ when $r\neq_E 1$. And from a philosophical perspective, the axioms of $\cal Q$ are supposed to deliver us an intended interpretation which associates indistinguishability to non-individuality. Such an intended interpretation concerns a metaphysical assumption about identity. If $s_1$ is indistinguishable from $s_2$ in a way such that $s_1\neq_E s_2$, then $s_1$ and $s_2$ are both devoid of identity. But what about the case where $s_1 \equiv_r s_2$ for $r\neq_E 1$? Where is the descriptive power of quasi-sets in order to cope with degrees of indistinguishability?

The point here is the evidence that quasi-set theory $\cal Q$ provides a limited view regarding indistinguishability in the quantum realm. That happens due to: (i) a clear separation between macro- and micro-atoms; (ii) a clear separation between the cases where indistinguishability entails extensional identity and those cases where that cannot happen; and (iii) the fact that $\cal Q$ was partially motivated by the Symmetrization Postulate as evidence for a lack of individuality among quantum particles. And the Symmetrization Postulate is a problem of combinatorics, while degrees of indistinguishability are expressed in the continuum of real numbers between 0 and 1. Such a limited vision seems to work just fine when we want to accommodate quantum statistics and quantum states of atoms (like Helium), since such phenomena may be reduced to purely corpuscular approaches (at least within the non-relativistic description of elementary particles). Nevertheless, indistinguishability in quantum theories may refer to quantum trajectories as well (i.e., field sources). And quantum trajectories are remarkably different from classical trajectories, due to restrictions imposed by the uncertainty principle. More than that, quantum trajectories behave quite differently from quantum particles as well, at least  in an experimental setup like a double-slit apparatus: at first sight it makes no sense at all to talk about gases of trajectories and it makes no sense to talk about trajectories permutations. In the particular case of the double-slit experiment we can find a striking example to illustrate the wave/particle duality in quantum mechanics. And that duality is not anticipated within $\cal Q$ axioms.

So, all of this leads us to a fundamental question: how is particle indistinguishability related to path indistinguishability in the double-slit experiment? After all, those paths in the double-slit experiment are simply field sources. How those degrees of indistinguishability between field sources are related to the indistinguishability between a single photon and a single photon in the double-slit experiment? A full account for this problem demands at least a field-theoretic approach for quantum phenomena, in order to establish relationships between photons and their sources, and photons and their detection. Nevertheless, it still remains the problem of determining what precisely is the meaning of a degree of indistinguishability, besides some number in the unit interval which is coincident with degree of coherence. A proposal for dealing with this problem is presented in the next Section. Final criticisms about quasi-sets are also provided.

\section{Discussion}

Quasi-set theory is a genuine logical approach to cope with collections of objects where a standard notion of identity seems to be irrelevant, like what it seems to happen in quantum mechanics and non-relativistic quantum field theory. Besides, quasi-set theory allows us a better understanding about the epistemological character of standard identity in ZFU \cite{Krause-05}. Nevertheless, the domain of explanatory power of quasi-set theory in the quantum realm seems to be limited to those quantum phenomena which are reducible to a purely corpuscular approach. Mandel's results concerning degrees of indistinguishability are physically measurable by means of fringes visibility in the double-slit experiment. And such degrees of indistinguishability cannot be understood within quasi-set theory from any relevant standpoint, whether in a mathematical or in a philosophical discussion about individuality. Quasi-sets introduce a weaker equivalence relation $\equiv$ called indistinguishability. If $x\equiv y$ but $x\neq_E y$, then it makes sense to say $x$ and $y$ are somehow devoid of identity. But what it means to say that two given trajectories in the double-slit experiment are indistinguishable with a degree $r\in [0,1]$? Is it sound to talk about some notion of indexed identity?

It may seem to be the case that those degrees of indistinguishability suggested by Mandel are a straight consequence of an allegedly lack of individuality of a single photon (if we assume a current metaphysical assumption inspired on Post, Schr\"odinger, Manin, French, Krause, and others). But how precisely would that happen? Well, a natural interpretation of Mandel's results is that fringe visibility $\cal V$ and which-way information $P_{ID}$ in the double-slit experiment are complementary quantities due to the wave/particle duality of light. That interpretation is consistent with the Uncertainty Principle proposed in \cite{Coles-14}. There is a variety of quantitative statements of the so-called wave-particle duality relations (WPDRs). For a list of references on this see \cite{Coles-14}. And in that same paper the authors show that WPDRs are equivalent to a modern formulation of the Uncertainty Principle. That means the apparent lack of individuality of a single photon cannot be properly modeled in mathematical terms without considering such intrinsic relations between particle and wave. The uncertainty principle proposed in \cite{Coles-14} can be seen as a generalization of Mandel's work. The authors' strategy in \cite{Coles-14} is to think of wave–particle duality in terms of guessing games, and anyone’s ability to win such games is quantified by entropic quantities. Specifically, an observer is asked to guess one of two complementary observables, namely, (i) which path the photon took in a Mach-Zehnder interferometer or (ii) which phase was applied inside the interferometer. The which-path (corpuscular aspect) and which-phase (wave aspect) observables are complementary. The Uncertainty Principle gives a fundamental restriction about which states the observer is unable to guess about both observables. That means the allegedly lack of individuality of a single photon is not enough for understanding Mandel's results, since there is a complementary principle between path indistinguishability and wave phase information. More than that, indistinguishability relations do not apply solely to quantum particles, but to photons sources (trajectories) as well.

Besides, other physical phenomena seem to be unreachable from a quasi-set-theoretic perspective, like those evidences of slight violation of the Symmetrization Postulate, as discussed above. In quantum field theory formalism, e.g., there are exotic symmetries which may be understood as a smooth interpolation between symmetric and antisymmetric states \cite{Modugno-00}. If we adopt the common philosophical vision that indistinguishability is directly related to permutability \cite{Saunders-06}, how can quasi-set theory deal with such interpolations and still provide a philosophical account for non-individuality? Are there measurable degrees of transition between symmetric and antisymmetric states? All those issues seem to point to a mathematics of the continuum rather than a discrete one. Arguably, quasi-set theory fails as providing philosophical grounds for the concept of non-individuality in quantum mechanics because there are no degrees of indistinguishability in $\cal Q$. And such degrees of indistinguishability seem to be related to interactions among quantum objects, as described by quantum (non-relativistic) field theories. That means such degrees of indistinguishability may have a great deal of impact on other physical phenomena besides interferometry, as suggested in \cite{Modugno-98}.

Another problem refers to the classical limit of quantum theories. A quite disturbing fact about the Symmetrization Postulate (stated in quantum mechanics and non-relativistic quantum field theories) is that it survives the classical limit \cite{Dieks-11}. Nevertheless, the (allegedly equivalent) Symmetrization Postulate stated in $\cal Q$ is a theorem (in that formal theory) only for the case of indistinguishable micro-atoms (if we are talking about two or more objects). Hence, if micro-atoms of $\cal Q$ are supposed to describe quantum objects, what do macro-atoms correspond to? Macro-atoms in $\cal Q$ are supposed to be individuals, while micro-atoms are not. So, $\cal Q$ seems to be a very limited choice for understanding the classical limit of quantum theories. Actually, according to Dieks and Lubberdink \cite{Dieks-11} it is quite possible that most discussions regarding the status of indistinguishable quantum particles are simply misguided. And that would happen because the Symmetrization Postulate refers to indices which do not correspond to any notion of particle at all.

So, what would be a way to cope with path indistinguishability (due to an apparent lack of individuality of photons) in the double-slit experiment and its relationship to coherence? If there is any answer to that problem, the same solution is supposed to cope with possible violations of the Symmetrization Postulate, and the classical limit of quantum mechanics.

As remarked before, the Symmetrization Postulate is simply some sort of vague intuition about indistinguishability which is hardly well definable (in the sense of providing some intuition about it) by means of a formal framework. Even in the case of quantum statistics, Bose-Einstein and Fermi-Dirac work fine only for ideal gases where interactions are not taken into account. And the Symmetrization Postulate is a mathematical statement within discrete mathematics. Nevertheless, Mandel's degrees of indistinguishability and related results concerning a new Uncertainty Principle \cite{Coles-14} tell us that we should think by means of a mathematics of the continuum (not a discrete one). So, one possible way to deal with all those problems is by considering the possibility there is nothing wrong with identity. Maybe the issue is not identity, but logic.

Within intuitionistic logic, for example, the double negation of a formula $F$ does not necessarily entail $F$. Smooth infinitesimal analysis \cite{Bell-08}, e.g., allows us to talk about multiple indistinguishable terms without ever considering either permutations or properties. In other words, there is no need for any fundamental axiom about the unobservability of permutations. And there is no need for considering Leibniz’s Principle of the Identity of Indiscernibles. That means we can precisely describe indistinguishable terms - which are not necessarily identical - and still avoid symmetrization as a postulate. That happens because ``$\neg (x\neq y) \Rightarrow x = y$'' is {\em not\/} a theorem in smooth infinitesimal analysis, since this well-known framework for differential and integral calculus is based on a intuitionistic framework with (intuitionistic) identity. Permutations can still be clearly defined, since this framework does not necessarily abandon identity. But permutations can be eventually indistinguishable in the sense that intuitionistic logic does not allow us to decide if certain collections $x$ and $y$ are either identical or different. In that case, $x$ and $y$ are indistinguishable.

Motivated by the Kochen-Specker Theorem, Isham and Butterfield \cite{Isham-98} provided a realistic interpretation for quantum mechanics based on intuitionistic logic. Their space of multiple semantic values is defined by a Heyting Algebra. Formulas from quantum mechanics are associated to such semantic values, which define a specific bounded lattice. Since every totally ordered set is a lattice, it is possible (at least in principle) to find out that Mandel's degrees of indistinguishability are simply semantic values expressed by elements from the real interval $[0,1]$, which is a particular case of a Heyting Algebra if $[0,1]$ is endowed with the usual $\leq$ relation. Thus, we do not need to consider quantum particles as non-individuals in any strong philosophical sense. An analogous situation happens with quantum trajectories in an experimental setup like the double-slit experiment. If $x$ and $y$ denote paths in a Mach-Zehnder interferometer, then the formula $x = y$ can be associated to a semantic value $r$ in a Heyting Algebra, if we admit the logic which is lurking in the quantum realm is an intuitionistic one. Eventually, $r$ can be a real number in the unit $[0,1]$ interval.

So, indistinguishability may not be any issue regarding identity. Indistinguishability may be a problem about decidability within a logical framework which is simply not classical. But that is a possibility that demands further investigation.

\section*{Acknowledgements}

The author states there is no conflict of interest.

I wish to express my deepest gratitude to D\'ecio Krause, Federico Holik, Jos\'e Acacio de Barros, Otávio Bueno and specially one anonymous referee of this journal for critical and insightful discussions concerning the contents of previous versions of this paper. The remaining problems in this paper are full responsibility of the author.

\end{document}